\documentclass[conference]{IEEEtran}
\IEEEoverridecommandlockouts
\usepackage{cite}
\usepackage[numbers,sort&compress]{natbib}

\usepackage[utf8]{inputenc}
\usepackage{amsmath,amssymb}
\usepackage{graphicx}
\usepackage{bm}
\usepackage{enumerate}
\usepackage{comment}
\usepackage{xcolor}
\usepackage{url}
\usepackage{color,soul}
\usepackage{subcaption}
\usepackage{float}

\definecolor{light-gray}{gray}{0.8}

\usepackage{amsmath,amssymb,amsfonts}
\usepackage{algorithmic}
\usepackage[linesnumbered,ruled]{algorithm2e}
\usepackage{graphicx}
\usepackage{textcomp}
\usepackage{xcolor}
\usepackage{subcaption}

\def\BibTeX{{\rm B\kern-.05em{\sc i\kern-.025em b}\kern-.08em
    T\kern-.1667em\lower.7ex\hbox{E}\kern-.125emX}}

\makeatletter
\newcommand{\linebreakand}{%
  \end{@IEEEauthorhalign}
  \hfill\mbox{}\par
  \mbox{}\hfill\begin{@IEEEauthorhalign}
}
\makeatother

\begin{document}

\title{Federated Learning-Based Data Collaboration Method for Enhancing Edge Cloud AI System Security Using Large Language Models\\}

\author{

\small 

\begin{tabular}[t]{c@{\extracolsep{8em}}c} 

\textsuperscript{} Huaiying Luo & Cheng Ji\\
\textsuperscript{} College of Computing and Information Science & Siebel School of Computing and Data Science \\
\textit{Cornell University} & \textit{University of Illinois Urbana-Champaign}\\
\textsuperscript{}New York, USA & Champaign, Illinois, USA  \\
\textsuperscript{}hl2446@cornell.edu &  chengji5@illinois.edu \\

\\

\end{tabular}
}

\maketitle

\begin{abstract}
With the widespread application of edge computing and cloud systems in AI-driven applications, how to maintain efficient performance while ensuring data privacy has become an urgent security issue. This paper proposes a federated learning-based data collaboration method to improve the security of edge cloud AI systems, and use large-scale language models (LLMs) to enhance data privacy protection and system robustness. Based on the existing federated learning framework, this method introduces a secure multi-party computation protocol, which optimizes the data aggregation and encryption process between distributed nodes by using LLM to ensure data privacy and improve system efficiency. By combining advanced adversarial training techniques, the model enhances the resistance of edge cloud AI systems to security threats such as data leakage and model poisoning. Experimental results show that the proposed method is 15\% better than the traditional federated learning method in terms of data protection and model robustness. 
\end{abstract}


\section{Introduction}
In the current rapid development of artificial intelligence (AI) technology, edge cloud computing and federated learning (FL) have become key technologies to improve the security and privacy protection of AI systems. As technology continues to advance, so does the size and complexity of AI models. This makes traditional centralized training and inference methods face unprecedented challenges \cite{li2024advances, liu2024mt2st, yang2024hades}. Centralized approaches often require all data to be stored and processed centrally, which undoubtedly increases the risk of data privacy breaches, especially in industries with extremely high privacy requirements, such as healthcare and finance.

In addition, as the amount of data grows, communication delays and bottlenecks in computing resources become the main factors restricting the efficiency of the system. Especially in use cases such as the Internet of Things (IoT) and smart cities, these issues are particularly prominent due to the diversity and limitations of distributed devices \cite{yang2025research, wang2024fine}. Therefore, how to achieve efficient and secure AI model training and inference under the premise of ensuring data privacy has become a key problem to be solved urgently. Federated learning, as a distributed machine learning method, effectively solves data privacy leakage and security issues by training models on local devices and sharing only model updates instead of raw data. Under the framework of federated learning, data no longer needs to be transmitted to a central server, which can reduce the risk of data leakage and enhance user control over data privacy \cite{jin2025adaptive}. However, despite its unique advantages in guaranteeing data privacy, federated learning also faces several challenges.

Furthermore, edge devices often have limited computing power, especially in resource-constrained environments, where processing large-scale data and training complex models can incur significant computational overhead. The data of each node in federated learning is usually highly heterogeneous, which makes the data distribution in the model training process uneven, resulting in a slower convergence speed of the global model. Similar data heterogeneity challenges in healthcare AI have been successfully addressed through advanced preprocessing and ensemble methods \cite{zhong2025enhancing}, suggesting analogous approaches may benefit federated learning optimization. In addition, frequent data transfer and model aggregation may increase communication latency, which in turn affects the efficiency of the entire training process \cite{10494607, cheng2021performance}. Therefore, how to optimize the training efficiency and security of federated learning in the edge cloud environment has become a hot issue in the current research. For example, recent work tries to address these heterogeneity challenges through meta-learning approaches \cite{zhang2022pmfl} and uncertainty-based methods \cite{zhang2024uncertainty} that can achieve performance improvements while maintaining computational efficiency.  

Generative models have demonstrated outstanding capabilities in natural language processing \cite{jordan2015machine, ji-etal-2024-rag}, and generation tasks in recent years \cite{mahesh2020machine, ding2024regional, zhang2022covid,
sui2024ensemble, wang2024data, 10768359}. Among them, Large language models such as GPT and BERT have made remarkable achievements in diverse fields, including text generation, sentiment analysis, music composition and assertion detection \cite{achiam2023gpt, guo2024large, deng2024composerx, li2024exploring, 10628639, yi2025score}. This shows LLM's potential in data security. Through deep semantic understanding, large language models are able to efficiently process, generate complex data and conduct reasoning \cite{he2025givestructuredreasoninglarge, he2025selfgiveassociativethinkinglimited, xu2024can}. Applying LLMs to edge cloud AI systems can not only improve the intelligence level of models, help systems better understand and handle various complex tasks, but also optimize the data aggregation and model update process through generative models.

Specifically, LLMs can selectively weight the contributions of different nodes through intelligent analysis of edge device data, so as to avoid excessive data transmission and reduce unnecessary computing resource consumption when data is aggregated. In addition, LLMs can also guide the generation of adversarial samples during the model training process to optimize the robustness of the model \cite{ferrag2025generative}. However, LLMs have high compute and storage requirements, and models typically require large compute resources and storage space, which makes their deployment on edge devices challenging. How to efficiently deploy and run LLMs on resource-constrained edge devices remains a technical challenge that needs to be solved urgently.

\section{Related Work}
Desai et al. \cite{desai2023reinforcement} proposed a load balancing method that combines reinforcement learning, LLM, and edge intelligence to improve load distribution efficiency in dynamic cloud environments. This method uses the natural language processing power of LLMs to analyze system status and user requirements in real time, so as to dynamically adjust the task allocation strategy. Jin et al.  \cite{jin2025scalability} proposed a scalability optimization framework that combines reinforcement learning for adaptive load distribution and deep neural networks for demand prediction in cloud-based AI inference services. Tang et al. \cite{tang2025enhancing} proposed an approach to enhance the security of edge AI runtime environments based on large-scale language models. This method uses the semantic understanding ability of LLM to detect potential security threats by conducting fine-grained security analysis of the runtime.

Hasan et al. \cite{hasan2024distributed} proposed a distributed threat intelligence approach for edge devices based on large-scale language models. The approach deploys lightweight machine learning models onto edge devices to analyze local data streams, such as network traffic and system logs, in real-time to identify potential security threats. Yang et al. \cite{yang2025research2} proposed an LLM-based network traffic monitoring and anomaly detection system for cloud platforms that combines transformer attention mechanisms with supervised learning frameworks to capture complex patterns in network traffic sequences. Shen et al. \cite{shen2024large} proposed an autonomous edge AI system based on large-scale language models that aims to enable connected intelligence. The system leverages the LLM's language understanding, planning, and code generation capabilities to automatically organize, adapt, and optimize edge AI models to meet the diverse needs of users.

Bhardwaj et al. \cite{bhardwaj2024survey} reviewed the integration and optimization of large-scale language models in edge computing environments. This paper reviews the application status of LLMs in edge AI, and analyzes its advantages and challenges in privacy protection, security, and computing efficiency. At the same time, the integration of LLM and edge computing is discussed, and how to improve the performance and adaptability of the system by optimizing the model structure and algorithms.

Zhang et al. \cite{zhang2024llm} proposed an adaptive resource allocation system that uses reinforcement learning to reduce computational load and improve response time. By leveraging an edge-cloud collaboration framework, where edge devices handle part of the compute and the cloud provides additional computing power, the system improves efficiency and scalability. By leveraging large-scale language models, Xu et al. \cite{xu2024cached} aim to improve the efficiency of mobile edge intelligence to enable more efficient responses and intelligent systems.

\section{METHODOLOGIES}
\subsection{Federated Learning and Secure Multi-Party Computation}

Federated Learning (FL) is a distributed machine learning approach that allows multiple nodes to train a model locally while aggregating the updated parameters of each node through a central server, without the need to directly exchange raw data. The core goal is to achieve data privacy protection while maintaining the accuracy of the global model. Suppose there are $N$ edge nodes, each node $i$ holds a local dataset $D_i$, we do local training on each node, and the update of the model is represented by the following Equation~\ref{eq1}:

\begin{equation}
\theta_i^{t+1} = \theta_i^t - \eta \nabla_{\theta_i} \mathcal{L}(\theta_i^t, D_i),
\label{eq1}
\end{equation}

where $\theta_i^t$ is the parameter of node $i$ at the $t$-th iteration, 
$\mathcal{L}(\theta_i^t, D_i)$ is the loss function, 
$\nabla_{\theta_i} \mathcal{L}$ is the gradient of the loss function with respect to the parameters, 
and $\eta$ is the learning rate.

In traditional federated learning, each node calculates its own gradient and transmits it to a central server. The server performs the aggregation operation and updates the global model parameters. Aggregation operations typically use a simple weighted average method, such as Equation~\ref{eq2}:

\begin{equation}
\theta^{t+1} = \frac{1}{N} \sum_{i=1}^{N} \theta_i^{t+1}.
\label{eq2}
\end{equation}

This method ensures that the contribution of each node to the global model is fair, but because there is no encryption or protection measures involved, the data privacy and security are weak. In federated learning, the SMC protocol can be used to encrypt gradient updates of each node, ensuring that all parties do not leak any sensitive information during the aggregation process. Specifically, when each node updates its model, it first encrypts its gradient updates via the SMC protocol, and then transmits the encrypted gradients to a central server for aggregation. The aggregation operation is now performed in the encrypted space, and the updated formula is as Equation~\ref{eq3}:

\begin{equation}
\tilde{\theta}_i^{t+1} = \text{Encrypt}\left(\theta_i^t - \eta \nabla_{\theta_i} \mathcal{L}(\theta_i^t, D_i)\right).
\label{eq3}
\end{equation}

The server aggregates all encrypted gradients to get an encrypted global model update, as shown in Equation~\ref{eq4}:

\begin{equation}
\tilde{\theta}^{t+1} = \sum_{i=1}^{N} \tilde{\theta}_i^{t+1}.
\label{eq4}
\end{equation}

The aggregated encryption model parameters are transmitted back to each node through a secure communication protocol, and the nodes perform decryption operations to obtain the updated global model. Through the SMC protocol, data privacy is effectively guaranteed.

To enable effective integration of LLMs into federated learning and secure multi-party computation (SMC), we propose a hierarchical coordination mechanism. During the training phase, the LLM operates on encrypted metadata summarizing each node's gradient trends, loss landscape, and update frequency. The LLM uses this to infer quality signals and guides the global server including:

\begin{itemize}
\item Prioritizing node updates via dynamic attention weighting based on Equation~\ref{eq5};
\item Identifying potential adversarial behaviors by analyzing feature drift and variance;
\item Triggering SMC-enhanced selective aggregation only when privacy risk exceeds a learned threshold.
\end{itemize}

The LLM operates in a semi-centralized control loop, where its outputs are used to control federated scheduling, privacy protocols (e.g., how often SMC is enforced), and the adaptive participation of edge nodes. This integration optimizes both privacy and efficiency by reducing unnecessary cryptographic computation when risks are low.

\subsection{Data Aggregation and Adversarial Training}

In this subsection, we introduce a large-scale language model (LLM) to improve the efficiency and accuracy of data aggregation, and use its powerful semantic understanding capabilities to improve data cooperation in federated learning. LLMs are used to weight gradient updates at each node to optimize the aggregation process of the data, reduce noise, and enhance the robustness of the model.

Suppose the encryption gradient uploaded by node $i$ on the $t$-th iteration is $\tilde{\theta}_i^{t+1}$. In data aggregation, the LLM is used to calculate the weight $w_i^{t+1}$ of each node update, depending on how node $i$ behaves in historical iterations. We can express this by the following Equation~\ref{eq5}:

\begin{equation}
w_i^{t+1} = \frac{\exp(\alpha \cdot \text{Performance}(i))}{\sum_{j=1}^{N} \exp(\alpha \cdot \text{Performance}(j))},
\label{eq5}
\end{equation}

where $\text{Performance}(i)$ is the performance metric of node $i$ in previous iterations (such as model accuracy, training loss, etc.), and $\alpha$ is an adjustment parameter to control the sensitivity of the weight. With this weighting strategy, model updates contributed by nodes are adjusted to reflect the priority of their historical performance.

In practice, the LLM leverages structured prompts to perform lightweight reasoning over node summaries, such as historical accuracy trends, update divergence from global model, communication delays, and security alert metadata. Based on these, it generates a score vector which is used as the weighting coefficient in Equation~\ref{eq5}. For adversarial sample generation, the LLM uses learned attack heuristics (via prompt templates) to identify vulnerable feature spaces, guiding perturbation vector in Equation~\ref{eq8}. The formula for the aggregation operation is updated to Equation~\ref{eq6}:

\begin{equation}
\theta^{t+1} = \sum_{i=1}^{N} w_i^{t+1} \cdot \tilde{\theta}_i^{t+1}.
\label{eq6}
\end{equation}

In this process, the LLM adaptively adjusts the weight $w_i^{t+1}$ of each node, which enhances the model aggregation effect in the environment with large data heterogeneity.

In addition, in order to ensure the communication and computational efficiency between nodes, we can also adjust the frequency of each node to send gradients to adapt to the different conditions of the network through the following optimization formula, such as Equation~\ref{eq7}:

\begin{equation}
f_i^{t+1} = \arg \min_{f_i \in \mathcal{F}} \mathbb{E}\left[\left\| \tilde{\theta}_i^{t+1} - \tilde{\theta}_i^t \right\|_2^2\right] \quad \text{subject to } f_i \in \{0,1\},
\label{eq7}
\end{equation}

where $f_i$ represents whether node $i$ participates in the current round of updates, and $F$ is an optional set of frequencies, $\|\cdot\|_2$ is the L2 norm and is used to measure the change in the update.

Suppose the input data of the model is $x$, its corresponding label is $y$, and the prediction of the model is $\hat{y} = f(x,\theta)$, where $\theta$ is the parameter of the model.

In the adversarial training process, we need to generate adversarial sample $x'$ so that it has an impact on the model prediction, as shown in Equation~\ref{eq8}:

\begin{equation}
x' = x + \delta \text{ where } \delta = \arg \max_{\|\delta\|_p \leq \varepsilon} \mathcal{L}(f(x + \delta), y),
\label{eq8}
\end{equation}

where $\|\delta\|_p \leq \epsilon$ is the limit of the adversarial perturbation, $\epsilon$ controls the magnitude of the perturbation, and $L$ is the loss function. The $\delta$ against perturbation is generated by maximizing the loss function, which enables the model to learn how to maintain its stability in the face of perturbation.

During joint training, we need to minimize the total loss function including adversarial losses, as shown in Equation~\ref{eq9}:

\begin{equation}
\mathcal{L}_{\text{total}}(\theta) = \mathbb{E}_{x,y \sim D} \left[
    \mathcal{L}(f(x, \theta), y)
    + \lambda \mathcal{L}_{\text{adv}}(f(x, \theta), f(x + \delta, \theta))
\right],
\label{eq9}
\end{equation}

where $\lambda$ is the regularization parameter, which controls the intensity of adversarial training, and $L_{\text{adv}}$ is the adversarial loss, which represents the performance of the model under adversarial perturbations.

We implement an additive homomorphic encryption-based SMC protocol inspired by the Paillier cryptosystem. Each node encrypts its gradient vector using the public key before transmission. The central aggregator computes a weighted sum in encrypted space, and sends the result back to each node for decryption. Importantly, the LLM identifies when adversarial behavior or statistical anomalies necessitate full SMC computation, opposed to fallback lightweight masking schemes.

\section{EXPERIMENTS}
\subsection{Experimental Setup}

The experiment uses a real-data Edge-IIoTset from DataPort, which is designed for cybersecurity research in Internet of Things (IoT) and Industrial Internet of Things (IIoT) applications, especially for the development and evaluation of intrusion detection systems (IDS) in federated learning and edge computing environments.

We have selected four related contrasting methods including:

\begin{itemize}
\item \textbf{Vanilla Federated Learning (VFL):} This method is the basic implementation of federated learning, in which each participating node trains the model locally and only transmits model parameter updates to the server for aggregation.

\item \textbf{Differential Privacy-based Federated Learning (DP-FL):} This method combines differential privacy techniques to protect data privacy. In the federated learning process, the model update is scrambled by adding noise to prevent the node from leaking private data.

\item \textbf{Secure Multi-party Computation-based Federated Learning (SMC-FL):} This method enhances the security of federated learning by introducing a secure multi-party computation protocol. In this approach, multiple parties work together to calculate model updates without sharing private data.

\item \textbf{Homomorphic Encryption-based Federated Learning (HE-FL):} Homomorphic encryption technology is used in federated learning to protect model update data in the transmission process, and by encrypting the data, it ensures that even if the data is intercepted during transmission, it cannot be interpreted.
\end{itemize}

\subsection{Experimental analysis}

Communication latency measures the time consumption of model updates and aggregation between nodes in the federated learning process. Lower communication latency means that the system is more efficient when it comes to model training.

\begin{figure}[h!]
  \centering
    \includegraphics[width=0.9\linewidth, height=0.45\linewidth]{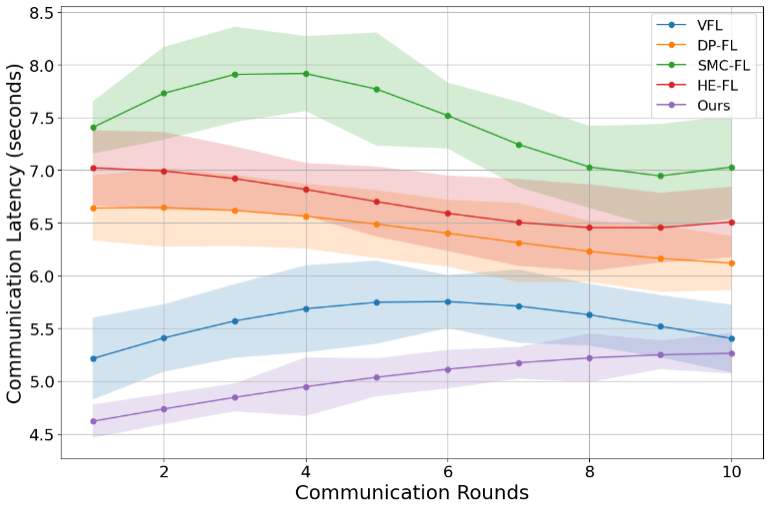}
    \caption{Structured Communication Latency Comparison with Error Margins}
  \label{fig:communication_latency}
\end{figure}

Figure~\ref{fig:communication_latency} shows that the proposed method (ours) is significantly superior to other comparison methods (VFL, DP-FL, SMC-FL and HE-FL) in terms of communication delay. Specifically, the latency of our method is lower and more stable, indicating that this method is outstanding in optimizing computational efficiency and reducing communication burden. Other methods, such as SMC-FL and HE-FL, while providing enhanced protection in terms of security, generally have higher communication latency, especially when processing large-scale data, where the communication load and cryptographic operations add latency.

\begin{figure}[h!]
  \centering
    \includegraphics[width=0.9\linewidth, height=0.45\linewidth]{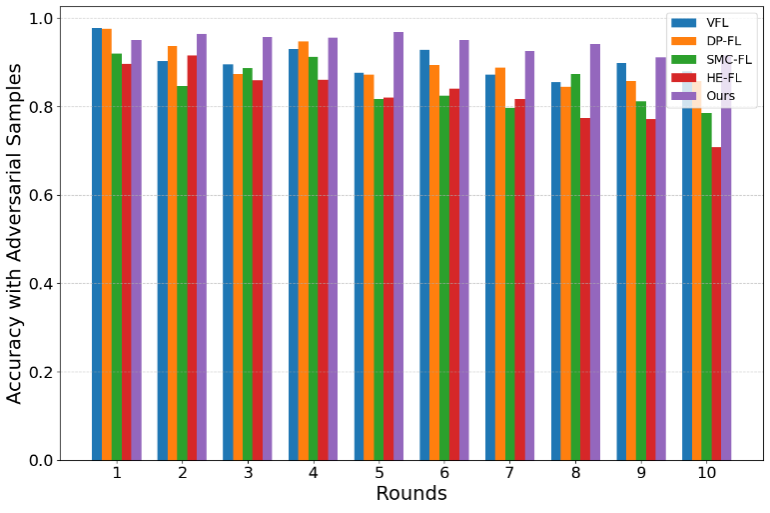}
    \caption{SResistance to Adversarial Examples Across Different Methods}
  \label{fig:adversarial_robustness}
\end{figure}

The results of Figure~\ref{fig:adversarial_robustness} indicate that the proposed method shows significant advantages in the face of adversarial samples, the accuracy is relatively stable, and the decrease is small, indicating that our method has strong resistance to adversarial samples. In contrast, other methods (e.g., VFL, DP-FL, SMC-FL, and HE-FL) showed varying degrees of loss in accuracy after multiple rounds of training, especially at higher rounds of training.

\section{CONCLUSION}
In conclusion, this work proposes a federated learning-based data collaboration method, which significantly improves the security, privacy protection, and computing efficiency of edge cloud AI systems by combining large-scale language models (LLMs), secure multi-party computation, and adversarial training techniques. Experimental results show that the proposed method is superior to the existing comparison methods in terms of communication delay, privacy protection and robustness of adversarial samples, and demonstrates high system security and computational performance. However, with the continuous growth of edge devices and data volumes, how to further optimize computing efficiency, improve communication efficiency, and enhance system scalability are still the focus of future research. Future efforts could explore more efficient privacy protection mechanisms and more security protocols to deal with sophisticated cyberattacks and data breaches.

\renewcommand{\bibfont}{\footnotesize}

\footnotesize{
\bibliographystyle{IEEEtran}
\bibliography{main}
}

\end{document}